\documentclass[]{aa}
\usepackage{natbib}
\usepackage{epsfig}

\usepackage{color}

\newcommand{\refe}{}
\newcommand{\refee}{}
\newcommand{\refeee}{}
\newcommand{\refeeee}{}

\def\GRS{GRS~$1915$+$105$}
\def\XTE{XTE~J$1550$-$564$}
\def\GRO{GRO~J$1655$-$40$}
\def\X1859{XTE J$1859$+$226$}

\begin{document}

\title{A possible \refeee{interpretation} for the \refee{apparent differences in} LFQPO types in microquasars}

\author{P. Varni\`ere\inst{1}
\and M. Tagger\inst{2}
\and J. Rodriguez\inst{3}} 

\institute{APC, AstroParticule \& Cosmologie, UMR 7164 CNRS/N2P3, 
Universit\'e Paris Diderot, CEA/Irfu, Observatoire de Paris, Sorbonne Paris Cite
10 rue Alice Domon et L«eonie Duquet, 75205 Paris Cedex 13, France.
varniere@apc.univ-paris7.fr
\and Laboratoire de Physique et Chimie de l'Environnement et de l'Espace, Universit\' e d'Orl\' eans/CNRS, Orl\'eans, France.
\and Laboratoire AIM, CEA/IRFU-CNRS/INSU-Universite Paris Diderot, 
CEA DSM/IRFU/SAp, Centre de Saclay,  F-91191 Gif-sur-Yvette, France.}

\date{Received <date> /
Accepted <date>}

\abstract{}
{In most microquasars, low-frequency quasi-periodic oscillations (LFQPO)  have been classified into three types (A, B and C 
\refee{depending on the peak distribution in the PDS and the shape of the noise)} but no explanation \refee{}  has been proposed yet.
The accretion-ejection instability (AEI) was presented in 1999 as a possible explanation for the \refee{fast varying} LFQPO that occur most often. 
Here we look at a possible generalization to \refee{explain the characteristics of}  \refeeee{the other two}  \refeee{LFQPO types.}}
{It was recently shown that when the disk approaches its last stable orbit, the AEI is markedly affected by relativistic effects. 
We focus on the characteristics of  the LFQPO that would result from the relativistic AEI and compare them with the different LFQPO types. }
{The effects of relativity on the AEI seem to be able to explain \refeee{most} of the \refee{characteristics} of the three types of LFQPO within one formalism. }
{}

\keywords{ X-rays: binaries,  
     stars: individual (GRS $1915$$+$$105$, XTE J$1550$$-$$564$, XTE J$1859$+$226$, GRO J$1655$-$40$, H$17143$-$322$), accretion disks}

\maketitle

\section{Introduction}

 Power density spectra (PDS) of black hole binaries show a high level of variability at all frequencies.    These PDS are usually fitted with the sum
of several Lorentzians. Depending on the spectral state of the black hole, thin features with
a high power are also often detected. By a widely accepted convention these thin features are referred to as QPOs if the value of 
the coherence ($Q =\nu_{\mathrm{centroid}}/FWHM $) is higher than $2$.
Here we are interested in the LFQPO s with frequencies typically in the range $0.1-20$Hz.
     These LFQPOs were originally considered to be one single phenomenon with varying frequencies.
    But during the $1998$ outburst of \XTE~, the LFQPOs displayed highly varying properties which led to classifying them 
    into three types labeled A, B and C  \citep[see for example][]{W99, R02}. 
    Since then, these three types of LFQPOs have been observed in several sources, including \GRS\  \citep{S07},
    indicating that they are three genuinely different types of QPO, and that \refeeee{mechanisms} common to all sources may give rise to them.

   Observational definitions of the three types of LFQPO are given in Table \ref{tab:lfqpo}, which is a summary of
   the results of \cite{R02}  and \cite{C04}, based on the microquasars \XTE\  \refeeee{and XTE J$1859$+$226$.}
\begin{table}[htbp]
\caption{Summary of the properties of the LFQPO types from \cite{R02}  and \cite{C04}, based on the microquasar \XTE.}    
\label{tab:lfqpo}      
\centering                                      
\begin{tabular}{l c c c}          
\hline\hline                        
Properties& type A & type B  & type C \\    
\hline                                  
frequency (Hz)  &   $\sim 6$  & $\sim 6$     & $0.1 - 20$    \\
amplitude (\%rms)    & $3-4$ & $\sim 4$ & $3-16$  \\
 Q  $\nu/FWHM$      &  $\sim 2-4$ &  $\sim 4$ & $> 10$ \\
phase lag (rad)    & $-0.6$ to $ -1.4$   & $0$ to $0.4$  & $0.05$ to $-0.4$  \\
\ \ subharmonic    & ...  & soft & soft  \\
\ \ first harmonic    &  soft & soft & hard  \\
coherence    &$< 0.5$    & $\sim 1$  & $\sim 0.9$ \\
HFQPO    &  4/4 & 6/9 & 5/51  \\
noise        & weak red & weak red & strong flat top  \\
\hline                                             
\end{tabular}
\end{table}   

   The type C LFQPO is the one that occur most regularly. \refe{It is observed in a \refee{spectral} state where the energy spectrum is close to a power-law
   	with a photon index of about $1.5$ to $1.9$ and with an exponential cutoff around $100$keV, namely the low/hard state}.
    \refee{Its main characteristic is a fast varying, highly coherent peak with a strong amplitude. It is observed simultaneously with a 
    strong flat top noise in the PDS.}
    It was proposed to be an expression of the accretion-ejection instability (AEI) in conditions where it dominates the inner region of 
    the disk \citep{TP99,RV02,V02}.
    
   On the other hand, types B and A tend to appear in softer states such as  the steep-power law state or  soft intermediate state, \refe{which shows 
 blackbody and power-law components of similar amplitude}. 
   \refee{The main peak frequency of the PDS is not varying much compared to type C and 
 is generally observed in absence of a flat top noise in the PDS. Among their differences, the type B typically} \refee{has
  a subharmonic as well as a first harmonic, while type A seems usually not accompanied by harmonics.
 Type B LFQPOs have a coherence similar to the type C, while type A are much less coherent.}
   A recent study of the \lq cathedral\rq\  LFQPO  of  \X1859 (categorized as a type B)  shows that the different peaks in the PDS 
   might not be related to one another  \refe{in just a simple harmonic manner} \citep{RV11}.
   There is also a link between the type of LFQPO and the presence of a high-frequency QPO. In Table \ref{tab:lfqpo} we see that HFQPO
   are almost always detected 
   \refe{when a type A or a type B LFQPO is also present, which is rarely the case when type C LFQPOs are present}.
   Furthermore, \cite{R02} showed that when both LFQPO and HFQPO are present, their Q-values appear anti-correlated, 
   which could hint at a competition mechanism between them. 
   
   Up to now, theoretical models of LFQPO have focused on type C LFQPO and tended to tie its frequency to a magnetoacoustical frequency
    \citep[e.g.][]{titar04}  or to the Keplerian frequency at some radius in the disk \citep[e.g.][]{TP99}. 
    The \refee{apparent} different behavior of the three types of
    LFQPOs reposes the question whether they are coming from different mechanisms or if it is the same mechanism, only expressing 
     differences
    occurring in the system. The first step in answering this question is to test if one mechanism can exhibit the different behaviors observed. 
   Here we focus on the  properties of the AEI and test if they could explain \refe{the peak distribution in} the three types of LFQPOs in a 
   single framework when relativistic effects are taken into account. 
   \refeee{To do this, we choose to focus on an interpretation for the peaks and not for the continuum of the PDS,} 
   \refe{  such as the band-limited noise (BLN) which sometimes occurs, or its correlations found with the 
   main peak's frequency. We consider this correlation to be an expression of their  origin and not an integral part of the QPO mechanism.}
           
     In section \ref{sec:AEI} we briefly review the properties of the AEI that made us consider it as a good candidate for the most common LFQPO, 
     the type C.
     We then investigate how relativistic effects can modify the properties of the AEI  \refe{when the inner edge of the disk approaches the last stable orbit.} 
     The following section presents  the LFQPO type that would result from the relativistic AEI (R-AEI).
     In the last section we turn to real data and see if the different flavors of the AEI can explain the observed characteristics of the three types of LFQPOs, first 
      \XTE, in which the different types of LFQPO have been most frequently observed. Subsequently,
     we consider several other objects to gain an exhaustive  view of the different types.

\section{The different flavors of the accretion-ejection instability}\label{sec:AEI}

\subsection{The AEI as a model for the common varying LFQPO}

The AEI was first introduced by  \cite{TP99}.  It is a global instability occurring in disks threaded 
by a poloidal magnetic field on the order of  equipartition \refee{with the gas pressure} (\lq fully magnetized\rq\  disk), 
namely when the plasma $\beta \sim 1$ and it  also requires
\begin{eqnarray}
\frac{\partial}{\partial r} \left( \frac{\kappa^2 \Sigma}{2\Omega B^2}\right) > 0,
\end{eqnarray}
\refee{where} $\Omega$ and $\kappa$ are the rotation and epicyclic frequencies (in a Keplerian  disk $\Omega = \kappa$), $\Sigma$ is the surface 
density and $B$ is the equilibrium magnetic field. This criterion is fulfilled  in disks with \lq reasonable\rq~density and magnetic field profiles. 
For the sake of completeness we present here the main properties of the AEI, as they have been discussed elsewhere 
\citep{TP99,CT01,VMTF03,VT02,RV02,V02,M09} in association with the type C LFQPO.

First of all, the AEI is an  instability, therefore it can grow naturally to a high amplitude without a need for an external excitation.
	It is also able to account for the following observational characteristics:

   {\tt -} The rotation frequency of the dominant $m=1$ mode, {\em i.e} the one-armed spiral
     predicted by the AEI is a few tenths of the Keplerian frequency at the inner edge of the disk. 
     This frequency is consistent with the LFQPO frequency \citep{TP99}.
		
   {\tt -} Linear theory and nonlinear simulations \citep{CT01} show that the AEI forms a standing wave pattern 
   that can saturate at a finite amplitude. \refee{It can thus account for the persistence of a QPO whose quality factor should be limited only 
   by the slow evolution of the factors (radial profiles of rotation, density, temperature, magnetic fieldÉ) that fix its frequency and amplitude, 
   or by nonlinear interaction between the modes when more than one is present.
	The result is therefore a pattern showing a fairly high coherence on  time scales that are long compared to the frequency/orbital period.
     The AEI naturally explains the thin features (QPO) seen in the PDS, while broad band components (e.g. the band-limited noise)
     would be either QPOs broadened by other effects or totally different phenomena. }

   {\tt -} Bbased on variations in disk properties at the location of the spiral wave, 
    the AEI partially reproduces the observed X-ray flux modulation \citep{VMTF03}.

   {\tt -} The AEI can transfer to the corona  a significant fraction of the energy and angular momentum that it extracts from the disk
   through Alfv\'en waves,
      thus providing a supply of energy that might feed the compact jet, often detected in the low-hard state, where the LFQPO is observed 
      \citep{VT02}.

   {\tt -}    Taking into account the effects of General Relativity through the existence of a last stable orbit and orbital velocity profile,
   when the inner edge of the disk approaches its last stable orbit,  
    we were able to explain the observed turnover in the correlation between the color radius 
    (related to the inner disk radius, as determined by the spectral fits) and the LFQPO frequency \citep{RV02,V02,M09}\\

\subsection{The relativistic AEI}      

        In a recent paper, \citet[][hereafter paper I]{V10} showed that in a disk \refee{whose}  inner radius \refee{is} close to the last stable orbit, 
         the AEI could co-exist with another instability, the Rossby wave instability (RWI),
        which was proposed \citep{TV06} as a possible explanation for the high-frequency QPO observed in microquasars. 
        This would therefore create a state with both  high-frequency and  low-frequency QPOs, just as is observed for the 
        type A and B LFQPOs in sources like  \XTE. This motivated us to investigat in more detail
         at the way the AEI is modified by relativistic effects, and how the observed  flux modulations would be affected.
  
\subsubsection{Range of interest for the location of the inner disk edge}
       
       For both instabilities (AEI and RWI) to coexist, we first need density and magnetic field profiles that meet the criteria \refe{Eq. 1}.
       Additionally, the AEI requires \refe{the magnetic field and the gas pressure to be close to equipartition},
      namely $\beta = 8\pi p/B^2 \sim 1$, and the RWI requires the inner edge of the disk to be close to the last stable orbit, $r_{in} \simeq r_{LSO}$.           
      It was previously shown that the frequency of the AEI alone is modified by relativistic effects 
      when the disk approaches its last stable orbit \citep{V02}. This modification causes a turnover in the correlation
      between the frequency and the color radius, which is compatible with  observations of \GRO\ and
      \GRS\  \citep[see][]{RV02, M09}.
      \begin{figure}[htbp]
\centering
\resizebox{\hsize}{!}{\includegraphics{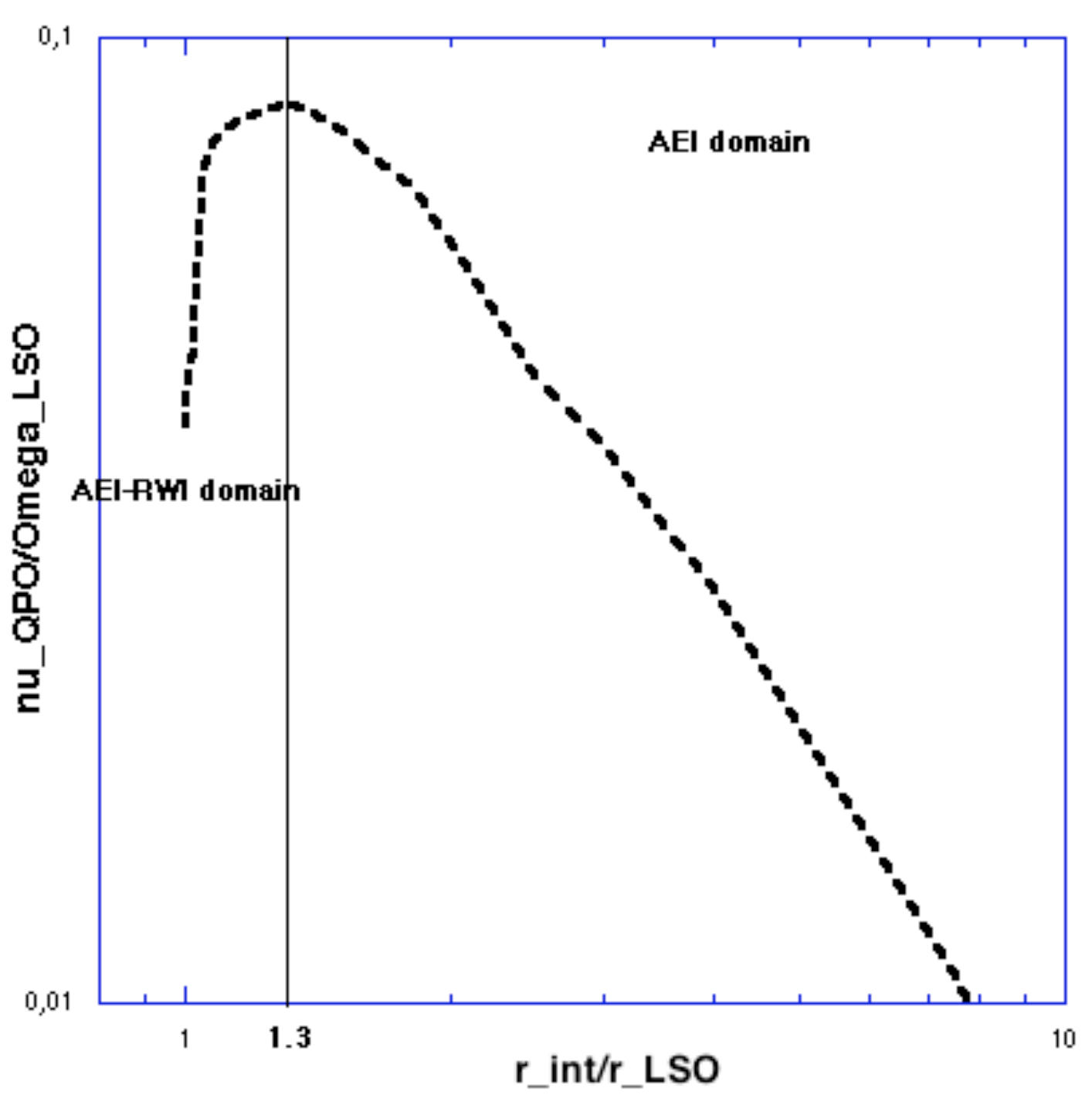}}
   \caption{{\small  Evolution of the frequency of the $m=1$ mode of the AEI as a function of the position of the inner edge of the disk with respect to
   the last stable orbit. The RWI occurs when the inner edge of the disk is inside $1.3\ r_{LSO}$. \refe{See \citet{V02} for more details.}
      }  }
\label{fig:AEI_RWI_domain}
\end{figure}
        Fig. \ref{fig:AEI_RWI_domain} shows the evolution of the frequency of the $m=1$ mode of the AEI as a function of the 
        position of the inner edge of the disk with respect to the last stable orbit.
       We see that the range of ratio $r_{in}/r_{LSO}$ in which both instabilities can coexist is the same in which  the AEI  would experience relativistic effects.
       In the right part of the curve in  Fig. \ref{fig:AEI_RWI_domain} the disk is mostly
       Keplerian, whereas in the left part relativistic effects near the inner disk edge modify the
       frequency-color radius correlation.
       Here we will focus on the top and left of the curve where we expect the relativistic
       AEI  to have most of its impact. We have shown \citep{RV02,M09} that the left of the curve can correspond, for instance, to 
       observations of  \GRO\  during its 1996 outburst,  \refee{ while observations of \GRS\  regularly  occupy both sides of the curve.}

\subsubsection{Numerical simulation results \refee{and mode spectrum}}
     
     In paper I we presented numerical simulations of a disk subject to  both  instabilities (the RWI 
     and the AEI) in its inner region, manifesting themselves as two spiral structures rotating at distinct frequencies.  
     \refe{Although their results may depend in an unknown manner on the limits} \refe{of \refee{the} numerical model and on our hypotheses, 
     they show quite stable features that we can compare with the observations.  }
     One of the main differences with simulations of the AEI alone is that instead of a dominating $m=1$ mode (where $m$ is the azimuthal wavenumber, 
     i.e. the number of arms of the spiral),  we tend to have higher-m modes dominating for both the RWI and the AEI.
     Indeed, in the presence of the $m=2$ mode of the RWI at the inner edge of the disk we tend to have  a dominating $m=2$ mode 
     of the AEI. These modes also tend to have a messier  Fourier representation than for
      the AEI alone \citep[see for example][for an example of the AEI alone]{CT01}. 
\newline
        
	Because we are interested in the distribution of peaks in the PDS (to compare with LFQPO observations), 
	we focus here on the evolution and strength of the $m=1$ to $4$ contributions in the simulation, regardless of  whether they are
	distinct modes or a mode and its harmonics. \refe{This does not allow us to compute the coherence of the different peaks,} 
	\refee{bur the AEI predicts quite a high coherence.} 
	 Up to now, when trying to explain the LFQPO, the focus was 
	on the strongest mode, here we deliberately looked at the evolution of the contribution up to the mode $m=4$.
	\refeee{Note that these modes are linearly independent. Their frequencies are eigenfrequencies of the system and depend on 
	global properties of the disk. Thus linear theory shows that their frequencies are close to, but not exactly in, harmonic relations.
	In a PDS they would therefore appear as close to a fundamental and several (sub) harmonics}.
	We are particularly interested in their relative strengths, which can be directly compared with observation.

\begin{figure}[htbp]
\centering
\resizebox{\hsize}{!}{\includegraphics{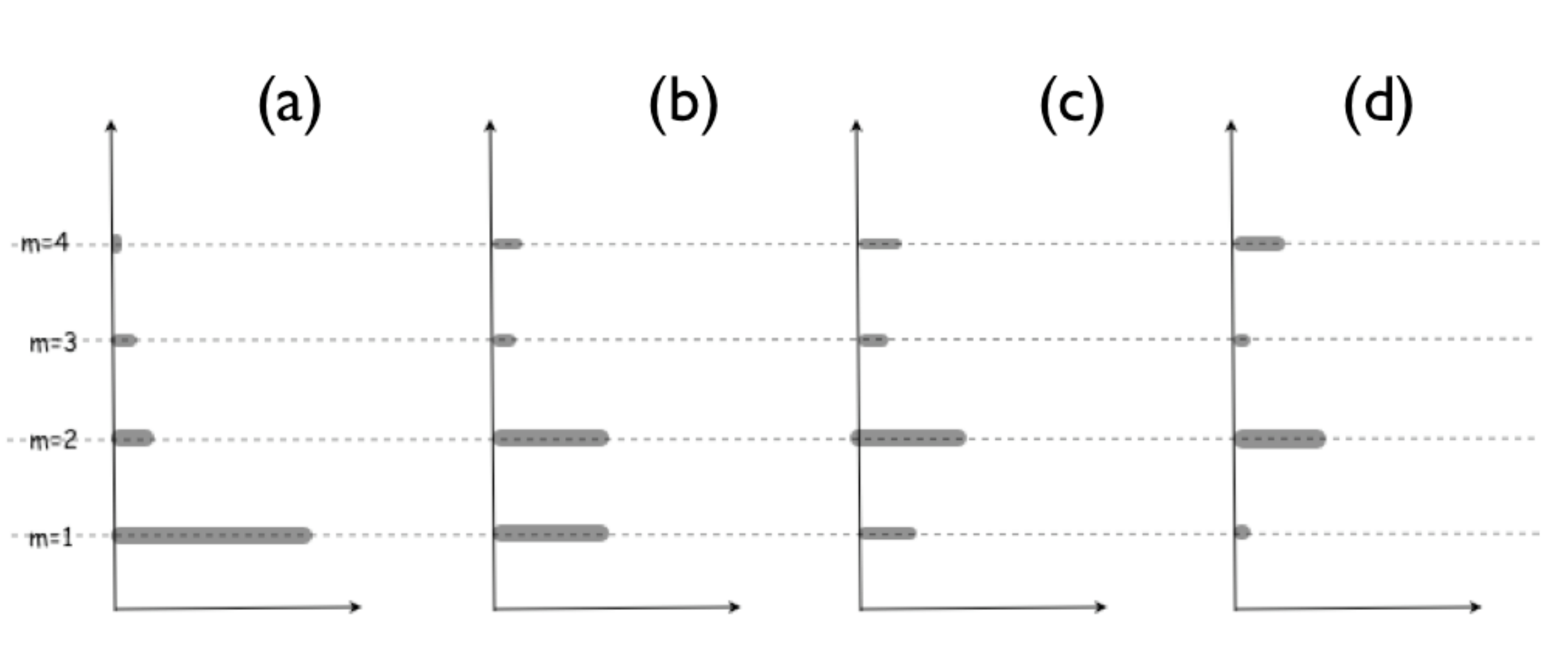}}
   \caption{{\small Shematic of the contribution behavior to the different mode of the AEI. (a) is a \lq standard\rq\ \refeee{peak distribution with higher modes of lower amplitude}, whereas (b) is  closer to the \lq cathedral\rq\ type.
   \refeee{The distribution in (c) has been observed in type C and B LFQPOs, while (d) is more characteristic of type A LFQPOs}.  }  }
\label{fig:3_types}
\end{figure}

	Fig. \ref{fig:3_types} is a schematic view of the strength of the first four modes in the different configurations we found in the simulation. 
	From left to right we see the evolution from a dominating $m=1$ mode to a strong couple of ($m=2,m=4$) modes. 
		
	In Fig. \ref{fig:3_types}a 
	we have a \lq standard\rq\  distribution of the mode strength:
 	the $m=1$ mode is the strongest, followed by a weaker $m=2$ mode,  and even weaker $m=3$ and $m=4$ modes. 
	\refe{This would correspond to the peak distribution of type C QPO that can typically be seen in the low-hard state} \refee{of \GRS\  and
	also at the beginning of several outburst sources such as \XTE\  and \X1859. } \\

	Fig. \ref{fig:3_types}b shows the case when the contribution of  the $m=2$ mode increases.  
	We arrive at a balance between the strength of the $m=1$ and the $m=2$ modes; 
	this would show a double peak in the PDS. In that case, the next mode in strength is the $m=4$ mode and not the $m=3$, 
	as it was before. \refe{ This corresponds to the specific case seen in XTE J$1859$+$226$ and dubbed cathedral QPO 
	\citep{C04,RV11}.}\\

	\refee{In Fig. \ref{fig:3_types}c  the contribution of the $m=2$ mode becomes dominant over the $m=1$ mode. In a PDS this would appear as if the stronger peak
	harbored a weaker subharmonic. The $m=3$ mode is weak  and of lower amplitude than the $m=4$ mode but 
	might be detected in bright sources.
	Depending on the PDS continuum and the shape of the noise, this would be labeled a type C (as in the later stage of outburst of \XTE\ and
	\X1859.) or a type B once the source has reached a softer state.}\\

	\refee{In Fig. \ref{fig:3_types}d  this effect is even more accentuated because $m=4$ becomes the second dominant mode after 
	$m=2$, while $m=1$ is weak 
	and  $m=3$ is barely detectable. This is similar to the type A QPO observed in outburst sources} \refeeee{(see  Fig.~\ref{fig:typeA} for a 
	fit of a type A LFQPO showing the two-peak structure)}.\\

\section{Characteristics of the R-AEI LFQPO and possible origin of the three types of LFQPOs}
    
    From these simulation results we see that the X-ray signal from the disk can be expected to be modulated in distinct manners, 
    well-suited to explain the observed differences between the three LFQPO types.
\newline
    
   {\tt - \refee{narrower} frequency range:} To show visible relativistic effects on the AEI,
   the inner edge of the disk needs to lie close to the last stable orbit, which in turn limits the range of frequency as
   depicted in Fig. \ref{fig:AEI_RWI_domain}. 
   Indeed, from linear analysis, \refee{ it was shown that relativistic effects start to appear  when the inner edge of the disk
   is as close as $3\ r_{LSO}$ \citep{V02}. When} the inner disk edge is within $1.3\ r_{LSO}$ the AEI  exhibit stronger relativistic changes,
      \refee{as shown on the  left} of the Fig.~\ref{fig:AEI_RWI_domain} 
   \refeee{This limited radial range also limits the frequency range to a factor of a few, i.e. much less than in the nonrelativistic case where 
   the inner edge of the disk can move much farther out, resulting in  frequency  variations by an order of magnitude.}
   \refee{The narrower range of frequencies in this region is compatible with the
	observed behavior of type A and type B QPOs, which we associate to the relativistic  flavor of the AEI.}  
   \refe{Indeed,  for \GRS\ the observed frequency on the left of the curve 
   in Fig.~\ref{fig:AEI_RWI_domain}  varies  between $\sim 6$ Hz and $\sim 9$ Hz \cite[see][for details]{M09}, 
   while for \XTE\ the variation for type B and A is observed to be between $5$ Hz and $9$ Hz \citep{R02}. 
\newline
   
   {\tt - presence of a  \lq sub\rq-harmonic:} For the classical AEI,  $m=1$ is the dominant mode, which means that
   the lowest frequency also has the highest amplitude.
    The other observed contributions  are much weaker, and given their weaker quality factor, will appear as a set of harmonics.
    For the R-AEI the dominant contribution is not  $m=1$ but a higher $m$ mode; \refee{for example
     the $m=2$ mode of the AEI is the dominant contribution in presence of the $m=2$ mode of the RWI. 
        In that case the weaker $m=1$ mode could be interpreted 
        as a subharmonic 
    of the dominant $m=2$ mode. This would be a much more preferable interpretation because classically it is very difficult for a 
    physical phenomenon to produce subharmonics.}
 \newline     
 
     {\tt - appearance related with the occurrence of HFQPO:} Within the R-AEI framework we see that the LFQPO 
     will be related to the \refeeee{HFQPO} because both instabilities have similar conditions (see paper I for more details).
 \newline

    \refe{Because of these characteristics the relativistic flavor of the AEI is a good candidate to explain the behavior of 
    some type C as well as most of type B and A, while the nonrelativistic AEI can explain the regular  type C 
    (by regular we mean without a subharmonic). 
    There is no one-to-one relation between our model of mode evolution and the definition of the three types of 
    LFQPO because we do not take into account the full PDS behavior, but only the LFQPO peak distribution. 
    \refeee{The phenomenological boundary between these three types takes into account other factors such as the behavior 
    of the continuum; these are not discussed here because they would require a full description of the consequences of the instabilities 
    on the disk, which is presently beyond our reach. At this stage, our interpretation thus only applies to and relies on the spectral 
    distribution and evolution of the QPOs.}}

\begin{description}    
   \item[\tt  type C LFQPO:]    \refe{Apparently, there are two cases for this type, \refeee{as one can see in the top row of Fig.~\ref{fig:typeCB}}.
   The most common one is the evolving LFQPO without a subharmonic:}
    this would be an expression of the AEI in its nonrelativistic form, {\em i.e.}  the dominant contribution comes from the $m=1$
    \refee{mode as seen in Fig.~\ref{fig:3_types}a}. 
    In that state the inner radius of the disk is far from its last stable orbit (larger than $\sim 3\  r_{LSO}$).
    \begin{figure*}[htbp]
\centering
\includegraphics[width=8.5cm]{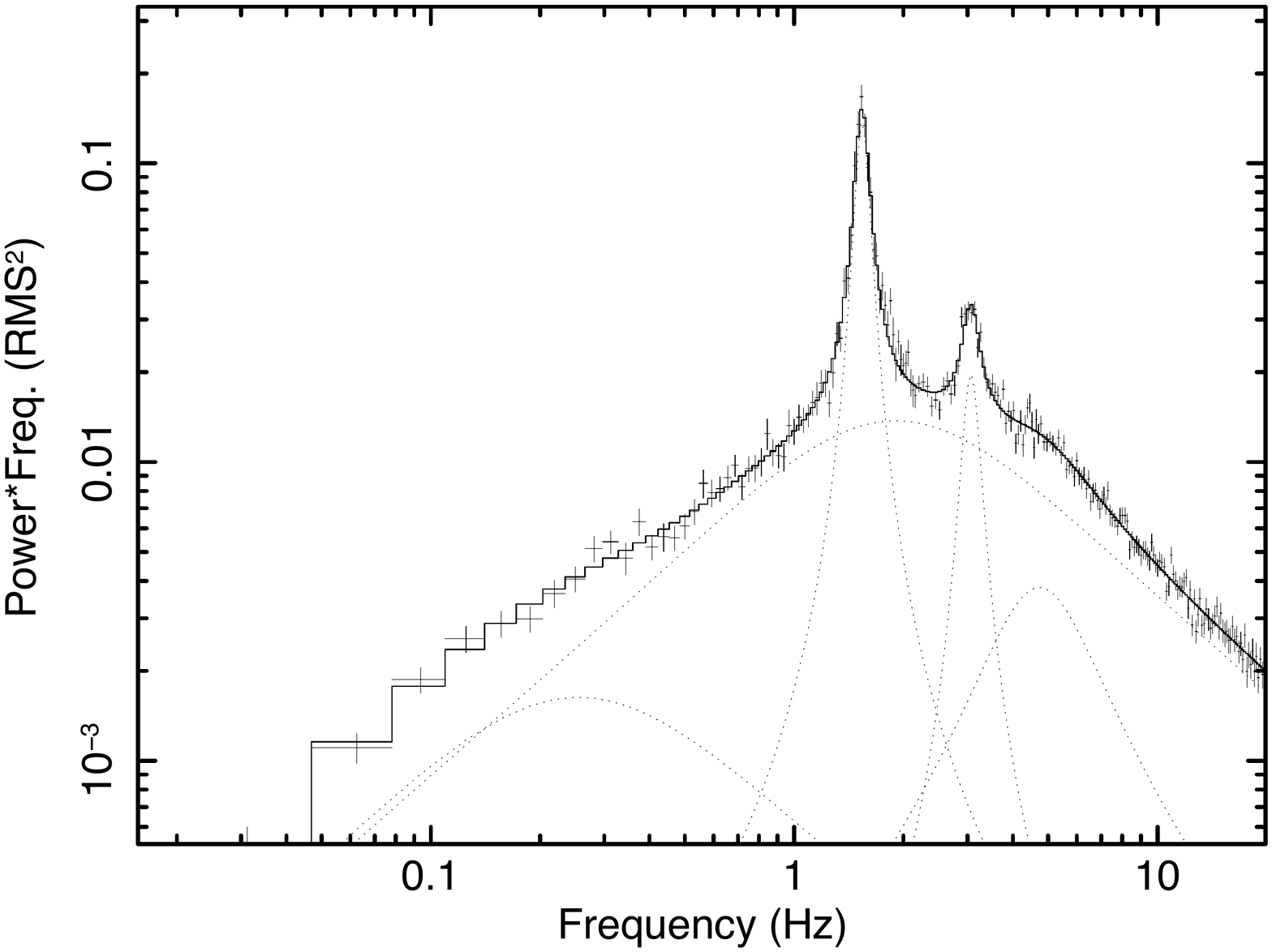}   \includegraphics[width=8.5cm]{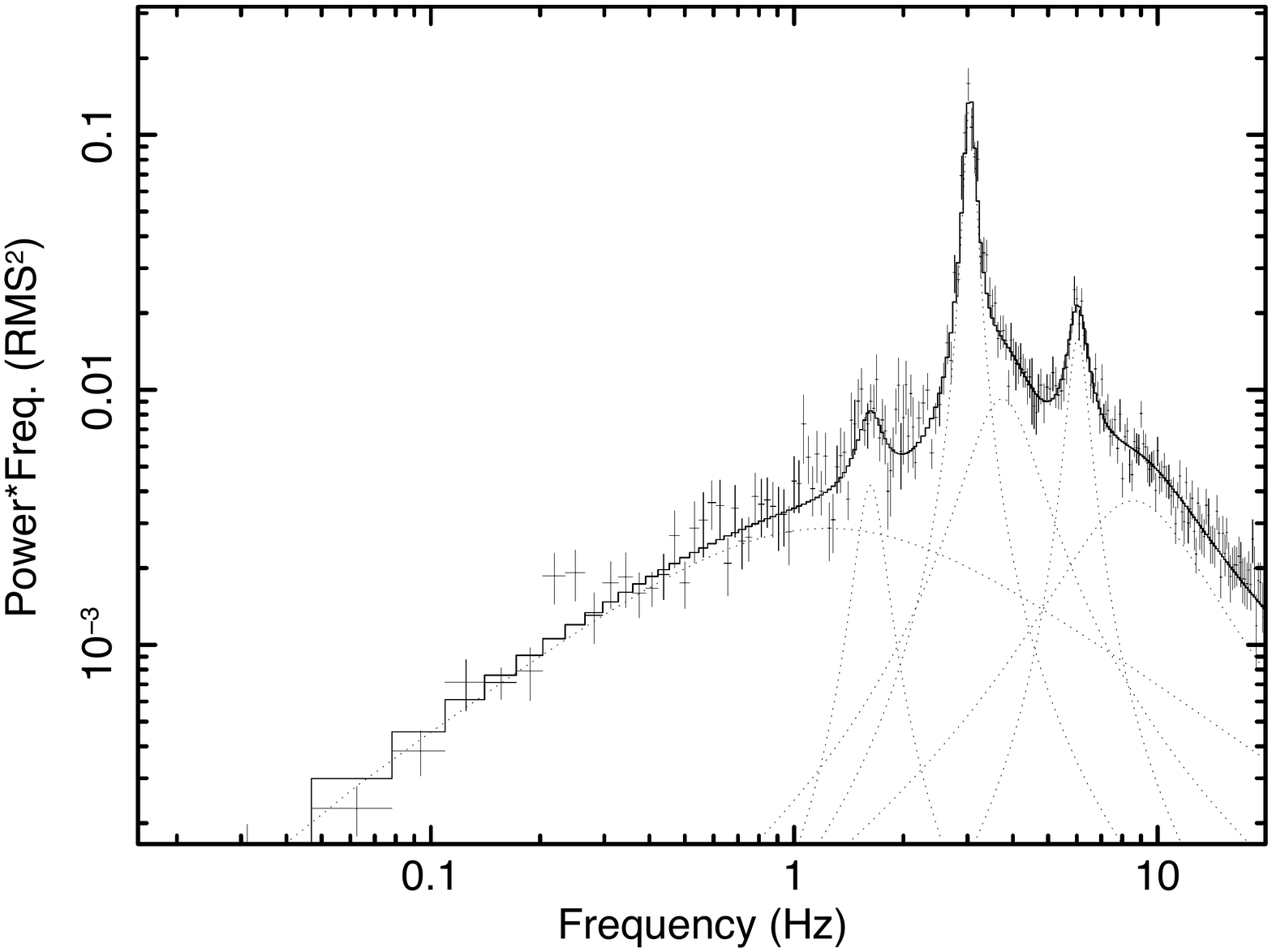} 
 \includegraphics[width=8.5cm]{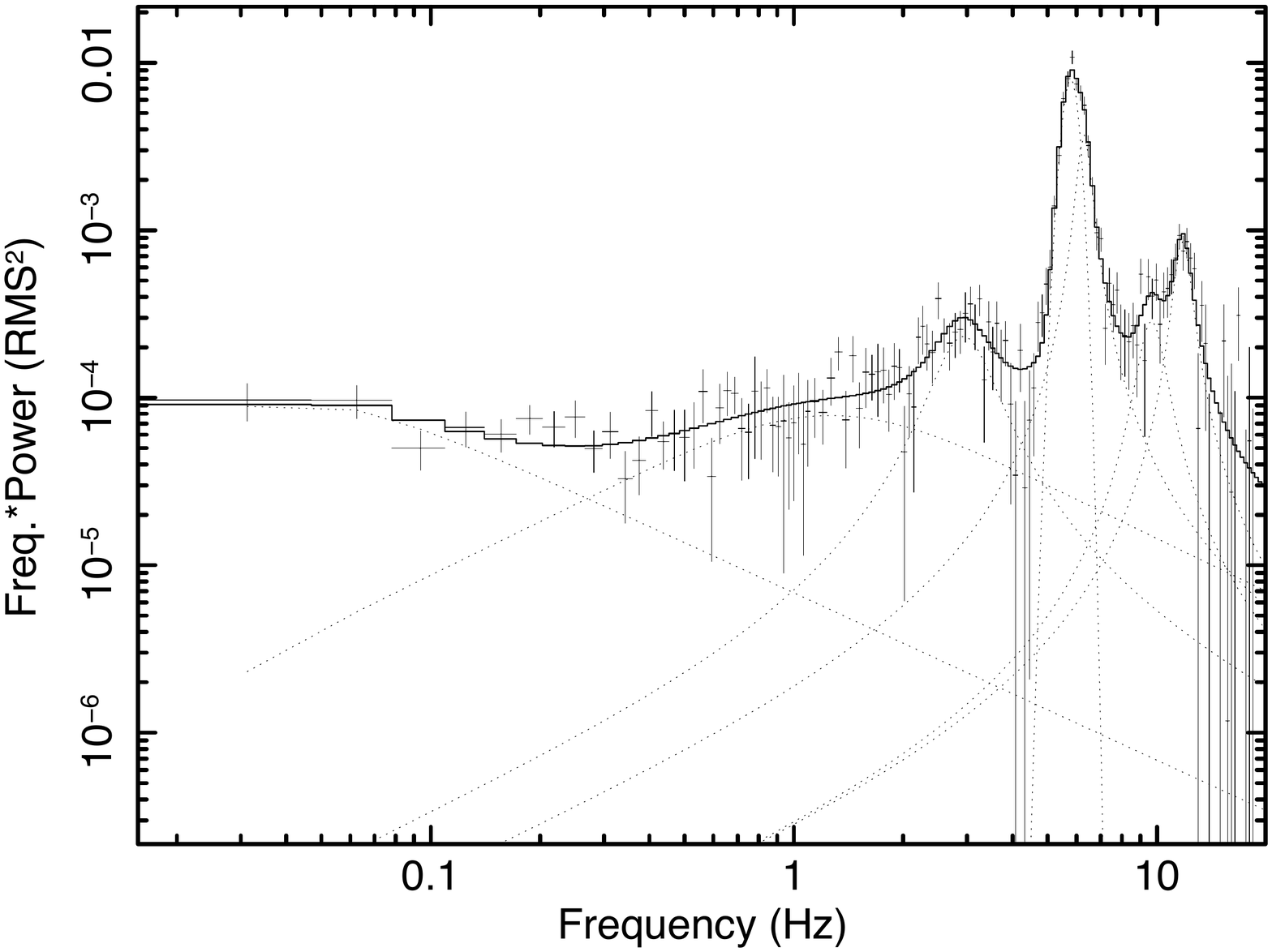}  \includegraphics[width=8.5cm]{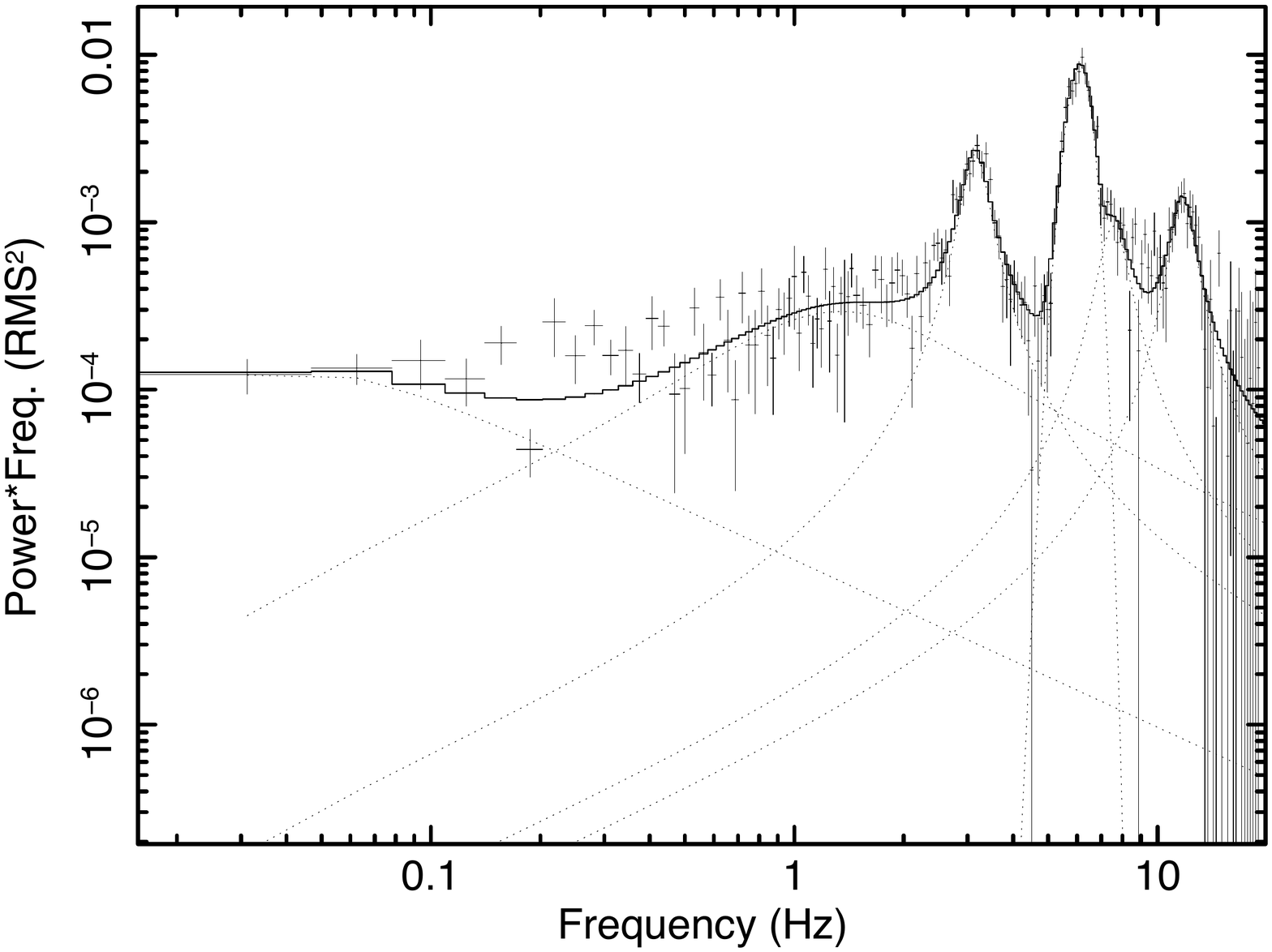} 
  \caption{{\small \refeee{Top: two different configurations of the type C LFQPOs.  Bottom: 
  two different configuration of the type B LFQPOs depending on the relative strength of the two 
   lowest peaks of the PDS taken during the 98 outburst of XTE J$1550$-$564$.}  }  }
\label{fig:typeCB}
\end{figure*}        
    \refe{But we also observe some type C LFQPOs, with subharmonics; we would explain them as an expression of the mildly relativistic-AEI 
     with a dominant $m=2$ mode. The main difference between this and type B  \refee{resides in the underlying disk, namely a different spectral state.}}
    \refee{Therefore, the type C are represented as  an evolution from case (a) to case (c) in Fig.~\ref{fig:3_types}}.

   \item[\tt type B LFQPO:]    this would be an expression of \refeee{a transition toward} the relativistic AEI-dominated state with the
      $m=2$ mode contribution dominating. 
        Because the $m=1$ mode is present but not dominant,  it would appear as a subharmonic of the dominant $m=2$ mode.
  	In that state the inner radius of the disk stays  close to the last stable orbit (within $1.3\ r_{LSO}$). 
	\refee{In Fig.~\ref{fig:3_types} type B would be either  case (b) or (c)}, \refeee{as one can see from the PDS of 
	XTE J$1550$-$564$ in the bottom row of Fig.~\ref{fig:typeCB}.}

   \item[\tt  type A LFQPO:]    this would be an expression of the R-AEI-dominated state with a strong contribution of both 
   the $m=2$ and $m=4$ modes,
\begin{figure}[htbp]
\centering
\resizebox{\hsize}{!}{\includegraphics{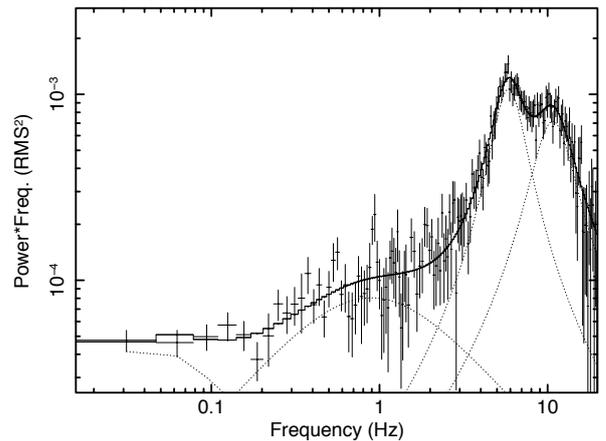}}
   \caption{{\small \refeee{Peak distribution of a type A LFQPOs taken during the 1998 outburst of XTE J$1550$-$564$.  }  }}
\label{fig:typeA}
\end{figure}    
    while the modes $m=1$ and $m=3$ are weaker. \refee{The case (d) of Fig.~\ref{fig:3_types} \refeee{and the   Fig.~\ref{fig:typeA}} 
    are examples of this}.
   Another important aspect of \refe{this case that we see in our simulations is that the modes are not as clearly 
   defined as in the previous cases.  This in turn would make the PDS appear \lq messy\rq\   with a broader width for the 
   detected QPO, \refeee{as one can see from the Fig.~\ref{fig:typeA}}.} 
\end{description}

\section{Comparison with observations}     
     
        We have shown that one instability was able to produce the main characteristics of the three types of LFQPO depending on the conditions
        in the accretion disk. 
       We now investigate the observations in  more detail, using several objects known to harbor different types of LFQPO
       to test \refeee{our interpretation}.

\subsection{The archetypical \XTE}     
 
       To compare the observational properties of the different types of LFQPOs with what is expected  from the 
       R-AEI, we look first at \XTE, which exhibits all different types with regularity.
       In this case type A and B LFQPO have a small range of frequency  \citep[of about $5$ Hz to $9$Hz, see e.g.][]{R02}, 
       which  agrees \refeee{with what we expect from the relativistic AEI.}

\subsubsection*{Subharmonic or dominant "first harmonic"}
  
    A striking feature of the LFQPOs during that outburst is the peak in the PDS at about half the frequency of the main peak.
    This feature was often noted as a \lq subharmonic\rq. In the framework of the R-AEI presented here, we would explain
    this feature by the nondominant $m=1$ mode.
    It is therefore important to determine if the dominant mode observed is possibly the $m=2$ mode or if is indeed the fundamental. 
\begin{figure}[htbp]
\centering
\resizebox{0.7\hsize}{!}{\includegraphics{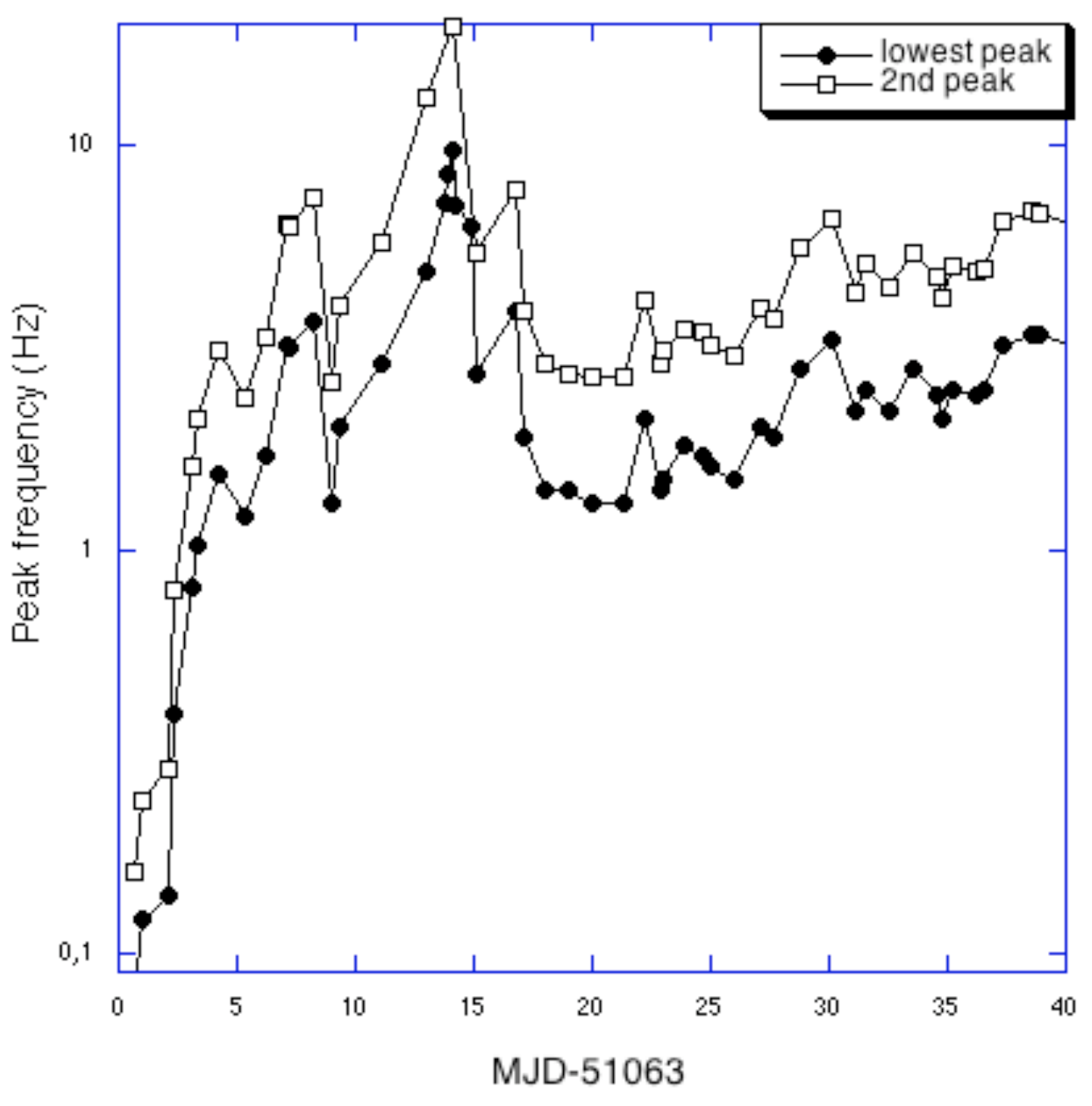}}
\caption{{\small Evolution of the frequencies of the two lowest peaks of the PDS  observed during the outburst of \XTE. 
\refe{Data taken from \cite{R02}.} 
}  }
\label{fig:2modes.1}
\end{figure}      

    To study the behavior of the two lowest peaks in the PDS, we took data from the 1998 outburst of \XTE\ as they are published in 
    \cite{R02}.
    Fig \ref{fig:2modes.1} shows  the evolution of the frequency of the two lower frequency peaks of the PDS 
    as functions of time.  
    The circles represent the lowest peak and the square the peak that is about twice (or more on a few occasions) 
    the frequency of the lowest peak. 
    The frequencies of the QPOs \refeee{are, within the errors, compatible with a harmonic relation.}

\begin{figure}[htbp]
\centering
\resizebox{0.7\hsize}{!}{\includegraphics{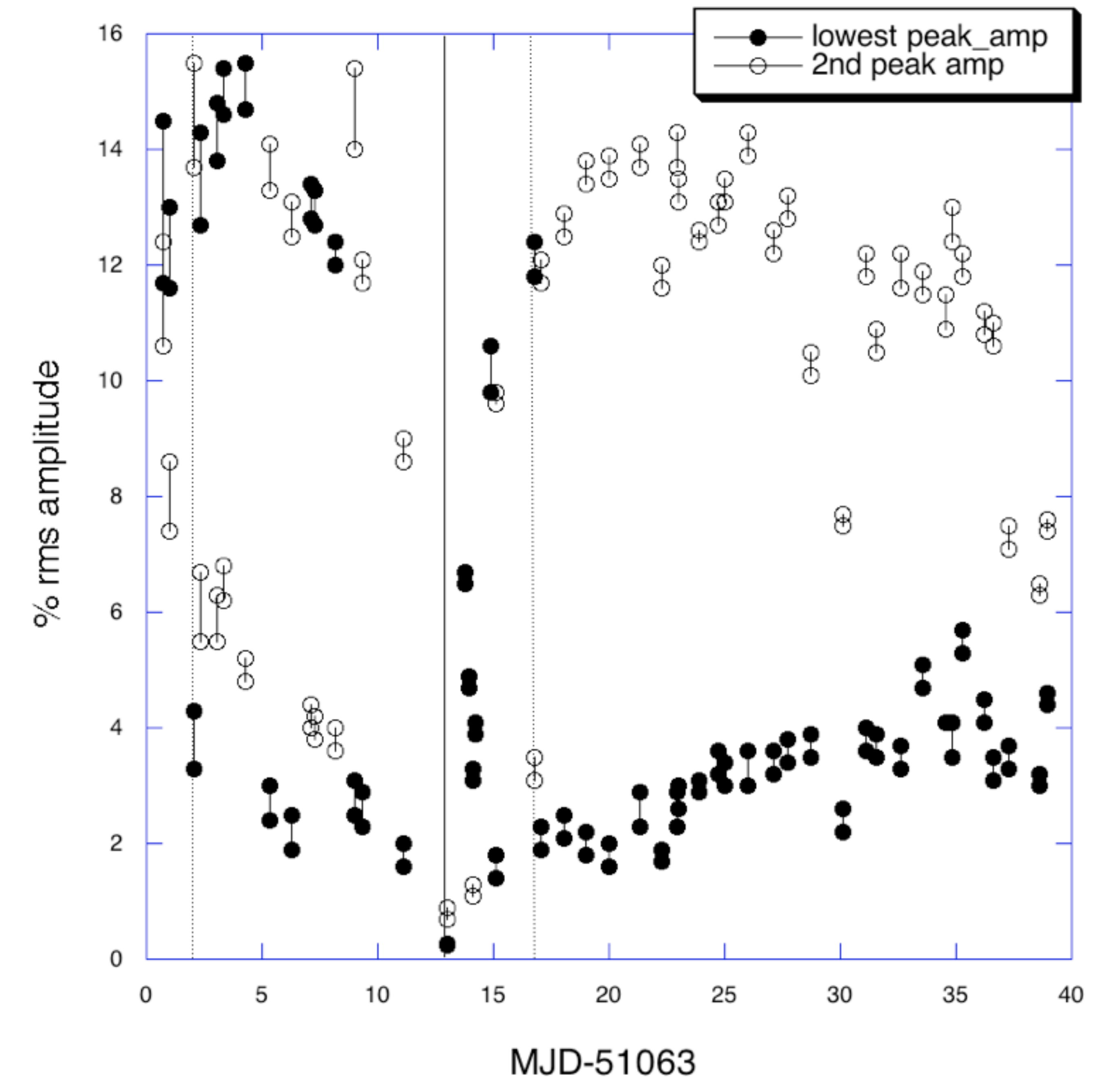}}
\caption{{\small Evolution of the rms amplitude of the two lowest peaks of the PDS.
 The circle represent the lowest peak at all times, while the square represents the peak detected at almost twice  the lower peak frequency. 
 \refe{ The vertical line represents the first detection of an HFQPO in the PDS. Data taken from \cite{R02}.} }  }
\label{fig:2modes.2}
\end{figure}      
       In Fig \ref{fig:2modes.2} we see that both peaks are present throughout the outburst, \refe{even though their relative strengths change.}
      This is coherent with an initially dominant contribution from the  $m=1$ mode, which is slowly replaced by the
      $m=2$ (and higher modes) as the disk approaches the last stable orbit and relativistic effects on the AEI becomes visible. 
      It is also interesting to note that around the time of the change of dominance for the different modes,  the  $\nu = 183$ Hz HFQPO 
      began to be detectable, reenforcing the idea that the disk is close to the last stable orbit and that relativistic effects should play a 
      role \refe{\citep{R02}}. 
 
\subsubsection*{Possible detection of the $m=3$ mode}
 
       To identify the first peak with the non-dominant $m=1$ mode, it would be interesting to look for a 
       rather weak
       $m=3$ mode that the model predicts.  
        Recently, \citet{Rao10} proposed that during this outburst up to four frequencies in the $1:2:3:4$ ratio were seen, even though the third 
        one is weak compared to the others. 
     
        During several observations of this outburst, it was required to add another peak  to the fit at about three times the frequency of the first peak, 
        as one can see in Fig. \ref{fig:m=3.1} \refeee{and in  the bottom right side of Fig.~\ref{fig:typeCB}.}
 \begin{figure}[htbp]
\resizebox{\hsize}{!}{\includegraphics{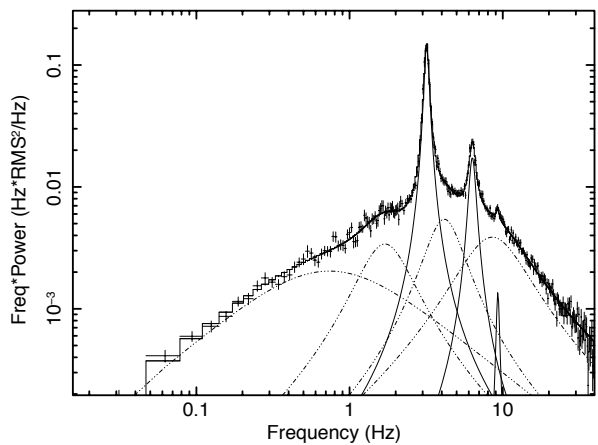}} 
  \caption{{\small  Observation of \XTE\ on MJD $51070.13$  one can see frequencies near the $1$:$2$:$3$ ratio.}  }
\label{fig:m=3.1}
\end{figure}  
       Figure  \ref{fig:m=3.1} is an observation on MJD $5107.13$, i.e. at the beginning of the outburst. 
       At that date the strongest peak is also the one with the lowest frequency and 
       we detect up to  $3\nu_o$  in the PDS. The fit requires seven Lorentzians ($\chi^2 =1.2$ for $248$ dof), 
       three of which are narrow features and  can truly be considered as QPOs,  and gives the following parameter for the
       peaks:\\
        Peak $1$ ($\nu=3.185 \pm 0.007$ Hz,  rms $=13.1_{-0.3}^{+0.4} \%$), \\
       Peak $2$ ($\nu=6.32 \pm 0.02$ Hz,  rms $=5.5_{-0.3}^{+0.2} \%$) and \\
       Peak $3$ ($\nu=9.29_{-0.09}^{+0.12}$ Hz, rms $=1.1 \pm 0.3 \%$).  \\
      Thus for that  observation we have significant peaks in the $1$:$2$:$3$ ratio in the PDS.  
              
 \begin{figure}[htbp]
\resizebox{\hsize}{!}{\includegraphics{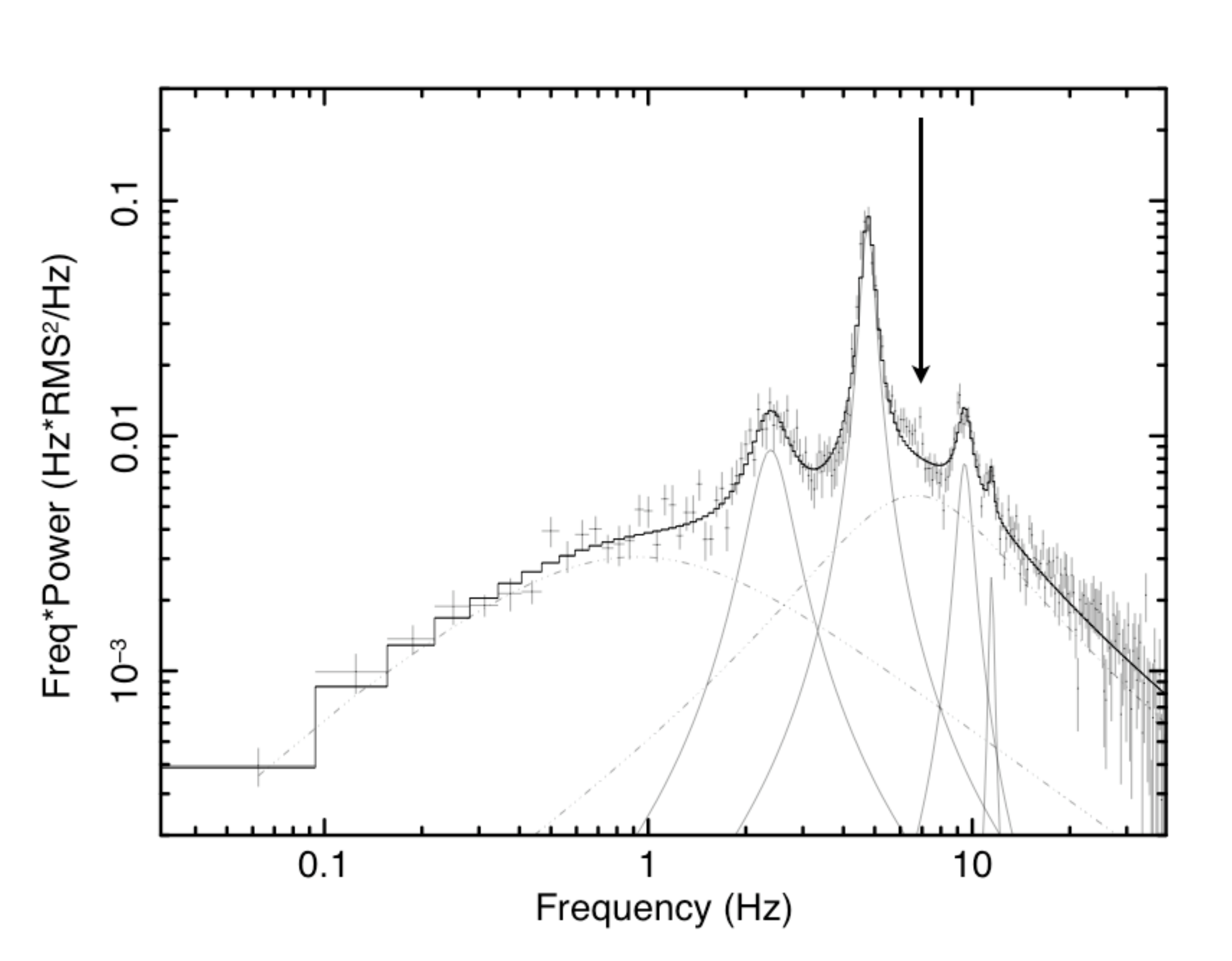}}
  \caption{{\small  Later during the outburst  of \XTE\ on  MJD $51099.6$  we see $1$:$2$:$4$ and a possible feature} where the $3$ would be. 
   \refee{The arrow indicates the position of this potential thin feature which is not accounted for in our fit.}}
\label{fig:m=3.2}
\end{figure}     
      On several other dates during that outburst  a \refe{feature} in the PDS around the $m=3$ frequency seems possible, 
      but is not required to obtain a good fit. Fig. \refe{\ref{fig:m=3.2} shows such an example; a thin feature, indicated by the arrow, may be present, 
      although it is not statistically required by the fit. Should this feature be a genuine one, it would then be} coherent with the presence
      of a weak $m=3$ mode that is clearly detectable when the $m=1$ mode dominates and becomes a weak feature as the $m=2$ mode 
      becomes dominant.

\subsubsection*{Link with the mode of the HFQPO}

        As was shown in paper I, the R-AEI  happens simultaneously with the the RWI, which was proposed to be at the origin of the HFQPO.
        This implies a strong link between the occurrence of type A and B LFQPO and the HFQPO.        
       As we showed in Fig. \ref{fig:2modes.2}, the dominant peak changed from the lowest frequency to the first harmonic on the same day that 
       the HFQPO occured.  
       \refe{It was also noted in \cite{R02} that the data suggest}
     \lq {\em that there is an anticorrelation between the amplitudes and coherence of HFQPOs and LFQPOs, respectively.}\rq\  
       \refe{This point \refee{is in favor of a link, and possibly a competition, } between the mechanism at the origin of both the HF and LFQPOs 
       of type B and A such as the one presented here.}

	When looking at the statistics of the occurrence of HFQPO with the different types of LFQPO, we see that
         the HFQPOs have frequencies  near $184$ Hz and $276$ Hz,  which appear while we have the
         type B (in six out of the nine observations of type B there was an HFQPO around $180$Hz)  and type A 
         (four out of the four observations of type A there was an HFQPO around $280$Hz) of LFQPO respectively. 
         
         Following our \refeee{interpretation} according to which the HFQPO is due to the RWI and the LFQPO is due to the AEI, we expect to sometimes
         see the $m=1$ mode of the RWI, especially during the \lq transition\rq\  from the mode $m=1$ to the mode $m=2$ of the
         AEI. 
         \refeee{This can be expected since the strength of either mode depends on many disk properties that affect their growth rates, 
         so that the usual dominance of the m=2 cannot be an aboslute rule.}
         During the 1998 outburst,  \refe{\citet{R02} detected at least one occurrence of an HFQPO at $92$Hz with a significance of $3.2\sigma$ and 
         an amplitude of $0.64 \pm 0.10\%$. A peak was detected at that frequency in several other observations but its Q value did not always qualify 
         it as a QPO.  The frequency of this peak is coherent with it being the \lq $m=1$\rq\ 
         associated with the $2$:$3$ ratio observed in that source}.
         This could be  the $m=1$ mode of the RWI instability associated with an LFQPO where the relativistic effects are becoming significant.
         Some more detailed study of this potential $m=1$ mode of the HFQPOs is required to properly compare it with the model. But this is still difficult
          at the moment because we have so few occurrences of this. 
            
\subsection{XTE J$1859$+$226$}   
 
      The outburst of XTE J$1859$+$226$ in 
      $1999$ was studied by \cite{C00}, who found HFQPOs. They also noted  a subharmonic
       to the LFQPO simultaneously with the HFQPO. This hinted toward different types of LFQPOs.
       Subsequently, \cite{C04} studied the same outburst, focusing on the timing. The detection of  the three types of LFQPOs was confirmed and showed
       similar  behavior to the one observed in XTE J$1550$-$564$.  Nevertheless, Casella {\em et al.} observed some interesting facts that were not seen
       in \XTE.
                
\subsubsection*{The lowest peak stays the same}

     \cite{C04} noted that during this outburst a subharmonic was always detected except for the first observations. 
     Interestingly, the frequency of the  only peak detected in the first observation is coherent with the frequency of the lowest peak 
     (which they called \lq subharmonic\rq) of the second observations.  
     This would suggest, in our model, that the $m=2$ mode became dominant early in the outburst.

      More importantly,  it was also noted that for the duration of the outburst  the \refeeee{lag related to the lowest peak is always
      negative, with a behavior independent of the type of LFQPO.} This suggests a common origin for the lowest peak in the PDS, 
     independently of the LFQPO type.  In our model, it would correspond to the $m=1$ mode of the AEI.

\subsubsection*{A possible feature in $1$:$3$ ratio with the lowest peak}    

	This outburst also exhibits case (b) of Fig. \ref{fig:3_types} that \cite{C04} called \lq cathedral\rq \  LFQPO (see Fig. \ref{fig:cathedral}), 
        where the two lowest peaks have similar amplitudes.  
        A complete timing analysis of the two observations showing this peculiar type is  presented in \cite{RV11}. 
       Here we use  Obs. 40124-01-24-00 to illustrate our model.
       The broadband 2.55--40 keV PDS was fitted with a sum of three broad and three narrow Lorentzians.
       \refe{A constant (in standard Leahy normalized units) was added to account for the white noise. This is represented by the linear line in 
       Fig. \ref{fig:cathedral} where the plot is in  power*frequency units. } The three thin features have the following parameters:\\
Peak 1: F= $2.94\pm0.02$ Hz, Q=$5.9$ , A= $2.8\pm0.1 \%$\\
Peak 2: F= $5.828\pm0.025$ Hz, Q= $7.3$ ; A=$4.7\pm0.1 \%$\\
Peak 3: F= $11.2_{-0.4}^{+0.3}$ Hz, Q= 9.5 , A=$1.1\pm0.3$ \% . \\
 \begin{figure}[htbp]
\hspace*{-0.5cm}
\resizebox{\hsize}{!}{\includegraphics{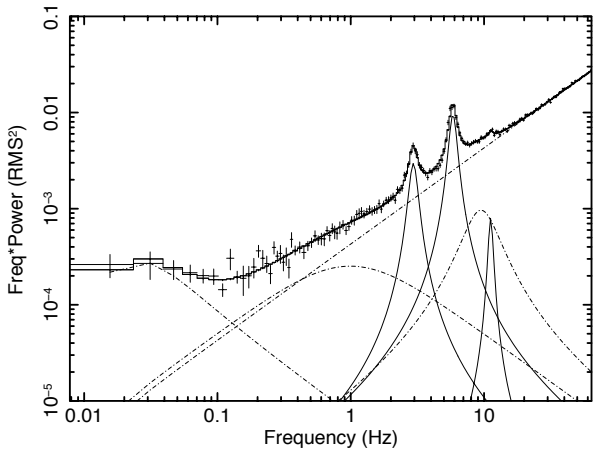}}
  \caption{{\small  Twin peaks of the cathedral subclass of LFQPO type B in XTE J$1859$+$226$ with an harmonic around $11$Hz
  and a broad feature around $9$Hz \citep{RV11}. }  }
\label{fig:cathedral}
\end{figure}  	
	The three peaks  therefore have frequencies that are compatible (at the $\sim 2\sigma$ level) with them being harmonics in the 
	ratio $1$:$2$:$4$ or $1/2$:$1$:$2$ \citep{RV11}.
	 \refe{Interestingly, there is \refeee{an additional} feature in the PDS at the expected frequency for an harmonic relation  
	 with peak 1 in a ratio $1:3$, 
	namely F=$9.0_{-0.5}^{0.4}$. This feature is broad (Q=1.5) but close to the QPO limit and has an amplitude of A=$3.2\pm 0.2 \%$.} 
	\refe{It is therefore possible that we are indeed observing the $m=3$ mode} along with the $m=1, 2, 4$. 
	XTE J$1859$+$226$ is therefore the second object (after \XTE) to show a feature in which the weak $m=3$ mode should be detected.
	Detecting this possible feature in another object strengthens its potential association with a weak $m=3$ mode,
	therefore putting the lowest peak in the PDS as the \lq fundamental\rq\  and not a subharmonic. 
     
\subsubsection*{Transition and competition between the types of LFQPOs}
          
       During that same outburst, \cite{C04} observed for the first time  
       several transitions between the different types of LFQPO within the same observation. They noticed the absence of 
      a direct transition from type A to type C LFQPO, which seems to emphasize the central role of  type B LFQPO 
      in the transitions.   
      At the same time, they also noted the apparent \lq balance\rq\  between the amplitude of the different peaks of the PDS. 
      This is particularly strong regarding the compared strength of the \lq subharmonic\rq\   and the "first harmonic". 
      Indeed, at the strongest rms amplitude of the "subharmonic" the \lq first harmonic\rq\   is barely detectable. 
      The reverse is also true.  This could be an expression of some kind of competition between the two. 
      
      This agrees well with our model, in which we associate each peak with a mode of the instability.
      This type of  \lq balance\rq\   is similar to the evolution seen between (c) and (d) of Fig \ref{fig:3_types}. 
      \refeee{Depending on the condition in the disk,  the dominant mode will change and therefore create 
      this transition.}\\

\subsection{Other objects in which type B can be inferred}    

       Using observations from \XTE\  and  \X1859,  we were able to find a good agreement between the characteristics of the R-AEI
       \refeee{and the observations of type A and B LFQPO}. 
       We now checked if  we can infer the presence of these types of LFQPO from the other characteristics
       shown here such as  HFQPO or the LFQPO lying in the left part of Fig \ref{fig:AEI_RWI_domain}.
               
\subsubsection*{GRO J$1655$-$40$}

     The LFQPO data of the 1996-97 outburst of GRO J$1655$-$40$ were not classified into the three types as is done for more recent observations.
     Nevertheless, this object seems a good candidate to harbor type B and A LFQPO based on several facts.
     First of all, not only HFQPOs were discovered in the source,  but the LFQPO and the spectral parameters have a 
     peculiar correlation when compared with other microquasars such as \XTE\  \citep{S00}. \refe{Indeed, the correlation between the LFQPO frequency 
     and the color radius is inverted compared to other sources.} 
     This behavior was later explained by using the AEI as a model for the LFQPO and requiring that the inner radius of the
     disk was close to its last stable orbit \citep{RV02,V02} during most of the outburst. Indeed,  GRO J$1655$-$40$ 
     was the first source to be discovered on the left side, namely $r_{int} < 1.3 r_{LSO}$, of Fig. \ref{fig:AEI_RWI_domain}.
         \refeee{This makes it, in our interpretation,  a manifestation of the R-AEI in the disk.}
        
     Looking more carefully at this outburst, we see that the LFQPO \refe{exhibits several of the characteristics of type B LFQPO. Indeed, it} 
     is shown to have an almost first-harmonic (\lq almost\rq because the two frequencies are not exactly 
     in an harmonic relationship but very close to it)  and sometimes that harmonic had a higher rms amplitude than the fundamental \citep{S00}.
      All of this make us consider  \GRO\ as another candidate in our search for the different types of LFQPOs. 
  	\refe{Indeed, we looked at several dates showing those characteristics (HFQPOs, two peaks in close harmonic relationship) and 
	confirm the presence of non-C LFQPOs in that source.}
     
\subsubsection*{GRS $1915$+$105$}
  
    A high-frequency QPO was discovered in \GRS~during the $\theta$ class of variability by \cite{B06}; more recently \cite{M09}
    showed that during the same class the color radius-frequency relation was different from the standard Keplerian one,
    and that the points were on the left side of the theoretical plot of Fig. \ref{fig:AEI_RWI_domain}.  It would be interesting to study the LFQPO   
    in more detail during the part of the observation where the LFQPO lies on the left of the correlation and see of what type they are.
    However, this  is difficult because the source is highly variable and spends little time on the left side of the theoretical 
    plot of Fig. \ref{fig:AEI_RWI_domain}.
    
    The search was made in other classes of the source by \cite{S07}, who found a transient type B LFQPO  in the classes $\beta$ and $\mu$. It
    would be interesting to look at the $\theta$ class where the source spends more time in the top/left side of the Fig. \ref{fig:AEI_RWI_domain}.
               
\subsection{H1743-322 and the search for HFQPO}       

     \refe{In the model presented here, the presence of type A and B LFQPOs happens in conjunction with HFQPOs. Observations tend to
     show a link. It therefore seems possible to use}
     the type of LFQPO as a tracer to narrow down the search for HFQPO in microquasars.

     \refe{This idea has} already been used observationally. 
     Indeed, in 2005 Homan \& Belloni  discovered a pair of HFQPOs in H$1743$-$322$ in a possible $3$:$2$ ratio and 
     mentioned the possible  presence of the \lq$1$\rq\  associated with the $3$:$2$ HFQPOs. 
     They also noted that, in presence of those HFQPOs,  the LFQPOs where different from the  \lq standard\rq \ C type and, 
     from their frequency, shape and harmonic content, found them to be closer to the A or B type as defined in \citet{W99}.  
     They also reported in one of their observations (labeled obs 2. in their paper)  what they interpreted as a transition from a type B to  A LFQPO,
     which would link the two types of LFQPO more closely. 
                   
	Subsequently,   \cite{R06} tried to detect more occurrences of these HFQPOs. Following the link that seems to exist between HFQPO 
	and the type of LFQPO, they decided to focus the search on observations that exhibit type B and A LFQPOs, namely 
	LFQPO without band-limited noise. \refe{To test the link between the different types of LFQPO and the presence of HFQPO,}
	 they used the \refe{shape of the PDS} to sort out the $130$ observations of that source
	\refe{in four groups}. The first group has no LFQPO, the second group exhibits   \lq standard\rq\  C type LFQPO with a band-limited-noise, and
	the two last groups had a non-C type LFQPO, the distinction between the two last groups being made depending on the flux level.
	They \refe{only} found HFQPOs in these two latter states, confirming a link between these non-type-C LFQPOs and HFQPOs.

\section{Conclusions}

     We have presented the \refeee{relativistic flavor of the AEI  as a possible cause for some of the distinct
     characteristics} for the type A and B LFQPOs in microquasars, \refeeee{with some of the semi-relativistic effects being visible already in 
     some type Cs.}
     \refeee{The AEI evolves from its Keplerian to its relativistic form when the fully magnetized disk extends close to 
     its last stable orbit. The effect is strongest when the disk extends down to its last stable orbit.}
     In these conditions two instabilities can occur, the AEI in its relativistic form, at the origin of the LFQPO, 
     and the RWI, at the origin of the HFQPO.
     The relativistic flavor of the AEI is a good candidate to explain the behavior of  type A and B of the LFQPO
     such as their \refeee{relatively} small amplitude in frequency, their (sub)harmonic content or the relationship with the HFQPO. 
    
     It can also be seen as a cue in the search for the elusive HFQPO. Indeed, the behavior of the LFQPO, 
     which is much easier
     to detect, can be taken as an indicator of the presence of HFQPO in the disk. In a first step, we were able to confirm that
     association by considering observations of \GRO\  during its previous outburst beforereturning to other objects.

\begin{acknowledgements}
The author thanks the anonymous referee that helped  clarify the paper to this final form.
This work has been financially supported by the GdR PCHE in France and the "campus spatial Paris Diderot".
JR acknowlegdes partial funding from the European Community's Seventh Framework Programme (FP7/2007-2013) under
grant agreement number ITN 215212 "Black Hole Universe".
\end{acknowledgements}

\end{document}